# Heat Sink Performance Analysis through Numerical Technique

B.Sri Aravindh
3rd Semester- Computer Science &Engineering
R.N.Shetty Institute of Technology
Bangalore

Dr.T.R.Gopalakrishnan Nair
Director - Research & Industry
D S Institutions
Bangalore
trgnair@yahoo.com

**ABSTRACT**
**The increase in dissipated power per unit area of electronic components sets higher demands on the performance of the heat sink. Also if we continue at our current rate of miniaturisation, laptops and other electronic devices can get heated up tremendously. Hence we require a better heat dissipating system to overcome the excess heat generating problem of using nanoelectronics, which is expected to power the next generation of computers. To handle the excessive and often unpredictable heating up of high performance electronic components like microprocessors, we need to predict the temperature profile of the heat sink used. This also helps us to select the best heat sink for the operating power range of any microprocessor. Understanding the temperature profile of a heat sink and a microprocessor helps us to handle its temperature efficiently for a range of loads. In this work, a method to estimate the normal response of a heat sink to various loads of a microprocessor is explained.**

## I. INTRODUCTION

We need to go far back in time to remember a CPU that was able to operate completely without a heat sink. The first Intel processors were already producing considerable amount of heat, but the low specifications allowed operation without any heat removal mechanism. A little later, as the processing speed increased, these processors required at least a passive heat sink for trouble free operation. However, for the last few years, as the processors got more and more powerful, it has become mandatory that a CPU requires a multifin heat sink as well as a fan that ensures reasonable air flow through the cooling fins as the overheated processors exhibit a shorter maximum life span and often results in problems like system freezes or crashes [2].

A heat sink is a device used in computers to remove the large amount of heat generated by components, including ICs such as CPUs, chipsets and graphic cards, during their operation. A heat sink is used to increase the surface area which dissipates the heat faster n keeps the ICs under safe operating temperature. Fans are also used to speed up this process. It usually consists of a base with one or more flat surfaces and an array of fin like protrusions to increase the heat sink's surface area contacting the air[2], and thus increasing the heat dissipating rate. A combination of a heat sink and a fan is widely used which maintains a larger temperature gradient by replacing warmed air more quickly. Heat sinks are made from good thermal conductors such as copper or aluminium alloy. Copper is significantly heavier and more expensive than aluminium but it is also roughly twice as efficient. The most common of a heat sink is a metal device (Cu or Al) with many fins. In this paper section 2 describes the heat sink types and its thermal resistance. Section 3 explains the least square model for estimating the parameters and predicting the best fit of the proposed heat sink in different loads of a microprocessor.

## II. HEAT SINK TYPES

The most common types of air-cooled head sinks [2] include,

- **Stamping:** Copper or aluminum sheet metals are stamped into desired shapes. They are the widely used type in air cooling of electronic components and offer a low cost solution to low density thermal problems.
- **Extrusions:** This method allows the formation of elaborate heat sink structures which are capable of dissipating more heat. As more number of fins can be made in this type, it increases the performance from 10 to 20%.

- **Fabricated/Bonded Fins:** The overall performance of an air-cooled heat sink can be improved significantly by increasing its surface area. This method allows us to bond fins to the aluminium base, hence increasing the surface area.
- **Castings:** This technology is used in high density pin fin heat sinks which provide maximum performance while using forced air system.
- **Folded Fins:** Corrugated sheet metal – either aluminum or copper, increases surface area, hence the performance. It is not suitable for high profile heat sinks, but it allows high performance heat sinks to be fabricated for specific applications.

## A. Heat Sink Thermal Resistance

The heat sink is designed in such a way that its thermal resistance is kept to the minimum possible value. The entire design process looks at the thermal resistance as the primary item to be calculated. The actual temperature of the microprocessor can be predicted with the calculated value of thermal resistance. Table 1 depicts the Thermal resistance of few electronic packages [1].

Table 1 Thermal Resistance for various electronic packages.

| Package | Junction to Case (°C/Watt) | Junction to Air (°C/Watt) |
|---|---|---|
| TO 3 | 5 | 60 |
| TO-39 | 12 | 140 |
| TO-220 | 3 | 62.5 |
| TO-220FB | 3 | 50 |
| TO-223 | 30.6 | 53 |
| TO-252 | 5 | 92 |
| TO-263 | 23.5 | 50 |
| D2PAK | 4 | 35 |

Power components have a maximum junction temperature, which must not be exceeded to prevent damage to the device. Devices are encapsulated in packages which have different levels of thermal resistance [2]. When designing power electronics, the heat dissipation of the device, coupled with any heat sinks, as well as the maximum power dissipated by the device, must be analyzed to insure that the device operates within allowable specified limits. For surface mount (SMT) parts, where the PCB copper is used as a heat sink, for 1 ounce copper the heat dissipation asymptotically approaches 1 square inch, in other words, having a PCB heat sink greater than 1 inch doesn't do you any good. There are some tricks that can help, such as placing vias, into the pad, so that heat is transferred to the bottom layer as wel [2]l. It's also possible to use surface mount heat sinks. Table 2 depicts the thermal resistance for the surface mount heat sinks with PCB copper [2].

Table 2 Thermal resistance for the surface mount heat sinks with PCB copper.

| Heat Sink | Thermal Resistance |
|---|---|
| 1 sq inch of 1 ounce PCB copper | 43 |
| 0.5 sq inch of 1 ounce PCB copper | 50 |
| 0.3 sq inch of 1 ounce PCB copper | 56 |
| Aavid Thermally, SMT heat sink | 14 |

The process of quantitatively estimating the trend of temperature rise in heat sink over a period of time can be done by using the method of Least squares, with which best curve fitting, can be obtained for various temperature levels of heat sink with different types of fans used in both high and low speed.

## III. LEAST SQUARE MODEL - CURVE FITTING AND REGRESSION

Field data is often accompanied by noise. Even though all control parameters (independent variables) remain constant, the resultant outcomes (dependent variables) vary. A process of quantitatively estimating the trend of the outcomes, also known as regression or curve fitting. The curve fitting process fits equations of approximating curves to the raw field data[2]. Nevertheless, for a given set of data, the fitting curves of a given type are generally not unique. Thus, a curve with a minimal deviation from all data points is desired. The best fitting curve can be obtained by the method of least squares.

The least squares parameter estimation method is a variation of the probability plotting methodology in which one mathematically fits a straight line to a set of points in an attempt to estimate the parameters. The method of least squares requires that a straight line be fitted to a set of data points such that the sum of the squares of the vertical deviations from the points to the line is minimized, if the regression is on Y, or the line be fitted to a set of data points such that the sum of the squares of the horizontal deviations from the points to the line is minimized, if the regression is on X. The regression on Y is not necessarily the same as the regression on X. The only time when the two regressions are the same (*i.e.* will yield the same equation for a line) is when the data lie perfectly on a line.

### A. The Least Squares method

The method of least squares is used to estimate parameters and to fit data. The best-fit curve of a given type is the curve that has the minimal sum of the deviations squared (least square error) from a given set of data. Suppose that the data points are $(x_1, y_1), (x_2, y_2), \ldots\ldots\ldots, (x_n, y_n)$ where x is the independent variable and y is the dependent variable. The fitting curve f(x) has the deviation error d from each data point, i.e.

$$d_1 = y_1 - f(x_1), d_2 = y_2 - f(x_2), \ldots\ldots d_n = y_n - f(x_n)$$

According to the method of least squares, the best fitting curve has the property ,

$$II = d_1^2 + d_2^2 + \ldots\ldots + d_n^2 = \sum_{i=1}^{n} d_i^2 = \sum_{i=1}^{n} [y_i - f(x_i)]^2 = \text{a minimum}$$

### B. Regression Model

Regression method is a statistical method for finding a line or curve ,the line of best fit that best represents a correspondence between two measured quantities .When the measurements are plotted as points on a graph and seem to fall near the same line, the least squares method may be used to determine the best-fitting line[1]. The method uses calculus techniques to find the minimum of the sum of the squares of the vertical distances of each data point from the proposed line. More generally, the process is called regression or, when the fitted curve is a line, linear regression. The best fit line associated with the n points $(x_1, y_1), (x_2, y_2), \ldots,$ $(x_n, y_n)$ has the form y = mx + b. where

$$Slope = m = \frac{n\left(\sum xy\right) - \left(\sum x\right)\left(\sum y\right)}{n\left(\sum x^2\right) - \left(\sum x\right)^2}$$

$$Intercept = b = \frac{\sum y - m\left(\sum x\right)}{n}$$

## IV. CASE STUDIES

For our study we have considered the surface mount heat sinks with PCB copper "1 sq inch of 1 ounce PCB copper" which has a medium level of thermal coefficient compared to the other same type heat sink as gven in the table 2. And Intel® Pentium® D Processor 915 microprocessor with a thermal coefficient of 63.4˚C for measuring the best fit of heat sink at various loads in different time and the "goodness of fit" of the least squares line, called the coefficient of correlation. The Coefficient of Correlation r is a number between -1 and 1. The closer it is to -1 or 1, the better the fit. For an exact fit, we would have r = -1 (for a negative slope line) or r = 1 (for a positive slope line). For a bad fit, we would have r close to 0. Intel® Pentium® D Processor 915 is the processor considered in our study for finding out the temperature variations at different time and load, to assess the performance of the heat sink used.

Table 3. Heat Sink Temperature Profile.

| Time in Seconds | Idle load-85W | Full load 150W | Fan Speed |
|---|---|---|---|
| | Temperature | Temperature | |
| Initial value | 20.2 | 20.2 | high |
| 5 | 20.2 | 22.4 | high |
| 10 | 20.4 | 25 | high |
| 15 | 20.5 | 27.2 | high |
| 20 | 21.2 | 29.6 | high |
| 25 | 21.8 | 32.5 | high |
| 30 | 22.0 | 33.8 | high |
| 35 | 23.8 | 42.6 | high |
| 40 | 25.9 | 54.7 | high |
| 45 | 26.4 | 55.2 | high |
| 50 | 27.5 | 56.0 | high |
| 55 | 28.0 | 56.5 | high |
| 60 | 28.4 | 56.8 | high |

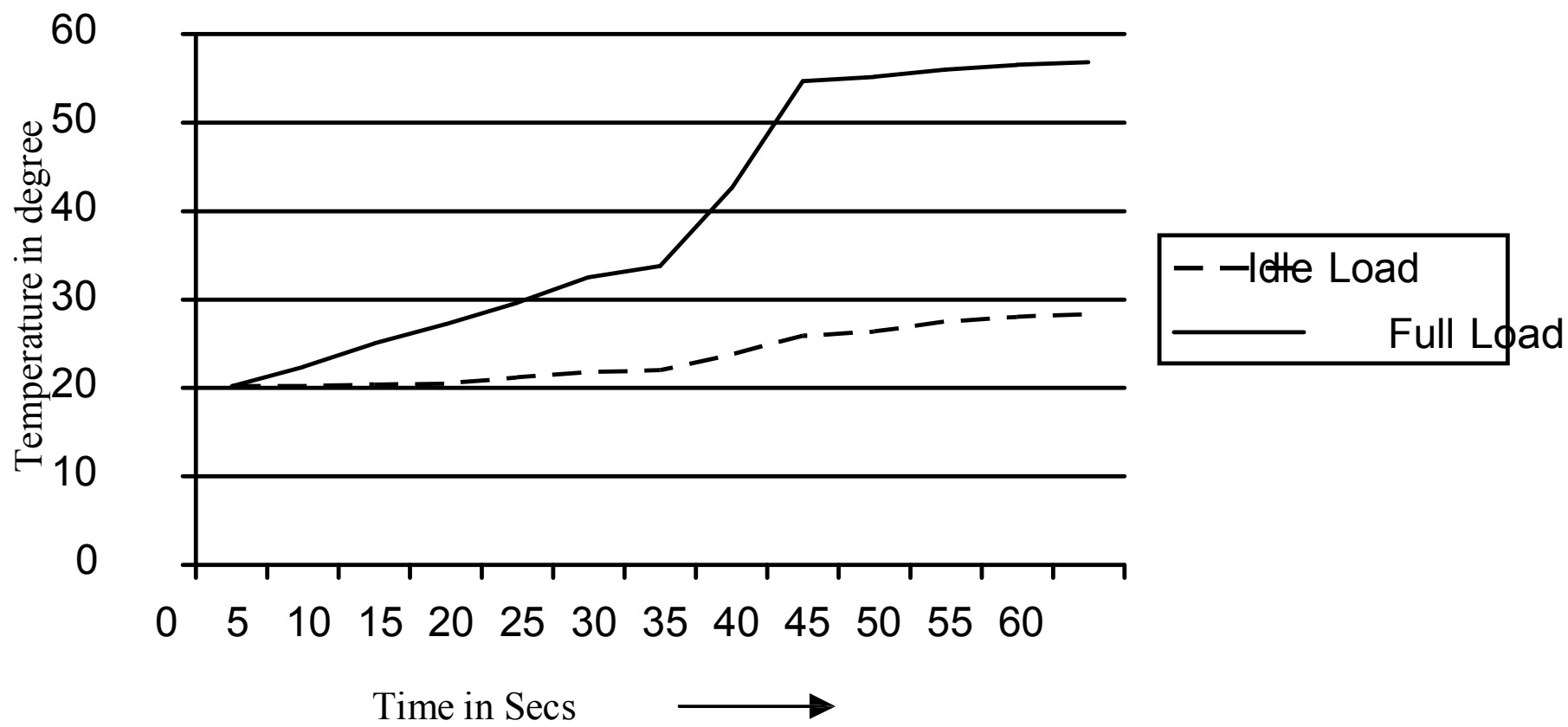

Figure 1 Heat sink temperature Profile

A. Functional Fit

The most frequent use of OLS was linear regression, which corresponds to the problem of finding a line or curve that best fits a set of data. In the standard formulation, a set of N pairs of observations $\{Y_i, X_i\}$ is used to find a function giving the value of the dependent variable (Y) from the values of an independent variable (X). With one variable and a function, the prediction is given by the equation Y = mx+b. The curve will be a straight line, a linear relationship between the two parameters x and y. Since we want almost linear curve we cannot minimize the vertical distances, but it is possible to minimize some reasonable combination of them [1]. Now, one reasonable combination of the distances would be their sum, but that turns out to be difficult to work with, because distances are measured in terms of absolute values. Instead, use of the sum of the squares of the distances where, no absolute values required. The line that minimizes this sum is called the best fit line, regression line, or least squares line, associated with the given data.

Table 4. Computed values of LS parameters.

| x | $y_1$ | $y_2$ | $xy_1$ | $xy_2$ | $x^2$ |
|---|---|---|---|---|---|
| Initial value | 20.2 | 20.2 | 20.2 | 20.2 | 1 |
| 5 | 20.2 | 22.4 | 101 | 112 | 25 |
| 10 | 20.4 | 25 | 204 | 250 | 100 |
| 15 | 20.5 | 27.2 | 307.5 | 408 | 225 |
| 20 | 21.2 | 29.6 | 424 | 592 | 400 |
| 25 | 21.8 | 32.5 | 545 | 812.5 | 625 |
| 30 | 22.0 | 33.8 | 660 | 1014 | 900 |
| 35 | 23.8 | 42.6 | 833 | 1491 | 1225 |
| 40 | 25.9 | 54.7 | 1036 | 2188 | 1600 |
| 45 | 26.4 | 55.2 | 1188 | 2484 | 2025 |
| 50 | 27.5 | 56.0 | 1375 | 2800 | 2500 |
| 55 | 28.0 | 56.5 | 1540 | 3107.5 | 3025 |
| 60 | 28.4 | 56.8 | 1704 | 3408 | 3600 |
| $\sum x = 390$ | $\sum y_1 = 306.3$ | $\sum y_2 = 512.5$ | $\sum xy_1 = 9937.7$ | $\sum xy_2 = 18{,}687.2$ | $\sum x^2 = 16{,}251$ |

$$slope = m_1 = \frac{n\left(\sum xy_1\right) - \left(\sum x\right)\left(\sum y_1\right)}{n\left(\sum x^2\right) - \left(\sum x\right)^2} = \frac{10(9937.7) - (390)(306.3)}{10(16{,}251) - 390^2} \approx 9289$$

$$Intercept = b_1 = \frac{\left(\sum y_1 - m\left(\sum x\right)\right)}{n} = \frac{306.3(1.9289)(390)}{10} \approx 105.857$$

$$slope = m_2 = \frac{n\left(\sum xy_2\right) - \left(\sum x\right)\left(\sum y_2\right)}{n\left(\sum x^2\right) - \left(\sum x\right)^2} = \frac{10(18687.2)\ (390)(512.5)}{10(16{,}251)\ 390^2} = 1.249$$

$$Intercept = b_2 = \frac{\left(\sum y_2 - m\left(\sum x\right)\right)}{n} = \frac{512.5(1.249)(390)}{10} = 99.6$$

The model is specified by an equation with free parameters. The values of the model parameters are being chosen to minimize the sum of the squared deviations of the data from the values predicted by the model. The Least Squares Method is widely used in building estimators and in regression analysis [1]. The computed data is used to construct an optimization function for the temperature control of the proposed heat sink. This technique gives temperature y, measured here by variation in time as a function of seconds, x. Here is a plot of y versus x.

Following is the formula for the best fit straight line. Least squares line is

$$y_1 = -1{,}9281x + 105.817, \text{ and } y_2 = -1.249x + 99.96.$$

We would like the temperature predicted by the best-fit line (predicted values) to be as close to the actual temperature at the selected time (observed values) as possible. It is possible, if the given points lie on a straight line,but in most of the cases the analysis shows that there will be some deviation, but there is a way of measuring the "goodness of fit" of the least squares line, called the coefficient of correlation. This is a number r between -1 and 1. The closer it is to -1 or 1, the better the fit. For an exact fit, we would have r = -1 for a negative slope line and r = 1 for a positive slope line. For a bad fit, r will be close to 0.

## V. CONCLUSIONS

Nowadays, the least square method is widely used to find or estimate the numerical values of the parameters to fit a function to a set of data and to characterize the statistical properties of estimates. It exists with several variations, Its simpler version is called ordinary least squares (OLS), a more sophisticated version is called weighted least squares (WLS), which often performs better than OLS because it can modulate the importance of each observation in the final solution. Recent variations of the least square method are alternating least squares (ALS) and partial least squares (PLS) [3]. When estimating the parameters of a nonlinear function with OLS or WLS the standard approach using derivatives is not always possible. In this case, iterative methods are very often used. These methods search in a stepwise fashion for the best values of the estimate. Often they proceed by using at each step a linear approximation of the function and refine this approximation by successive corrections [6]. The techniques involved are known as gradient descent and Gauss-Newton approximations. They correspond to nonlinear least squares approximation in numerical analysis and nonlinear regression in statistics. Neural Networks constitute a popular recent application of these techniques.

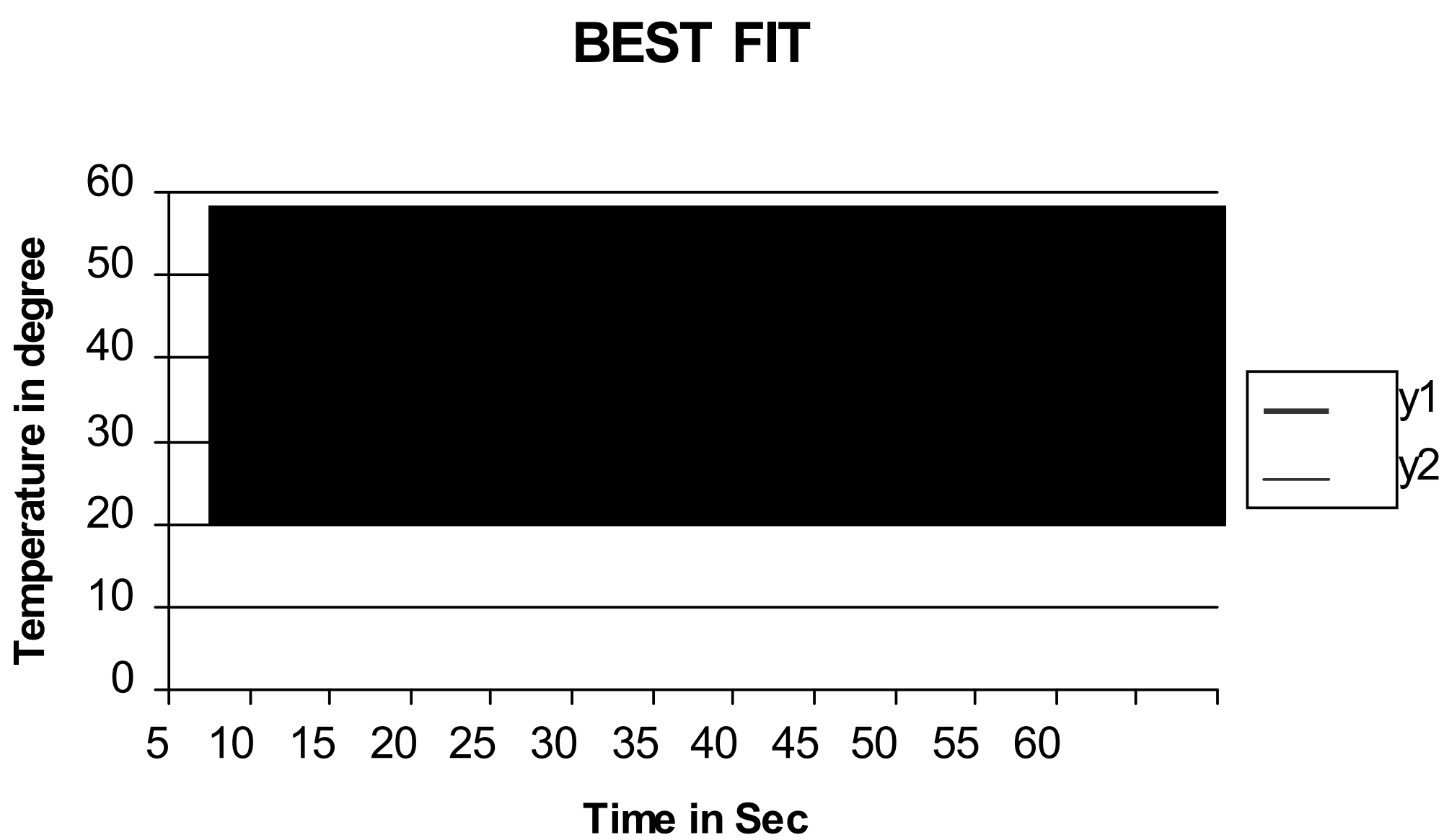

Figure 2 Best Fit Curve